\documentclass[10pt,conference]{IEEEtran}
\usepackage{setspace}
\usepackage[margin=0.5in]{geometry}
\usepackage{amsmath}
\usepackage{amssymb}
\usepackage[dvips]{graphicx}
\usepackage{color}
\usepackage[tight]{subfigure}
\usepackage{bm}
\usepackage{hyperref} 
\usepackage{enumerate}
\usepackage{booktabs}
\title{Distribution System Load and Forecast Model}

\author{
    \IEEEauthorblockN{Raffi Avo Sevlian\IEEEauthorrefmark{1}, Siddarth Patel\IEEEauthorrefmark{2} and Ram Rajagopal\IEEEauthorrefmark{2}\\}
    \IEEEauthorblockA{\IEEEauthorrefmark{1} Department of Electrical Engineering, Stanford University\\}
    \IEEEauthorblockA{\IEEEauthorrefmark{2} Department of Civil and Environmental Engineering, Stanford University}
}

\begin{document}
\maketitle

\begin{abstract}  
This short document provides experimental evidence for the set of assumptions on the mean load and forecast errors made in \cite{Sevlian2014A_Outage} and \cite{Sevlian2014B_Outage}.  
We show that the mean load at any given node is distributed normally, where we compute the mean and variance.
We then present an aggregation-error curve for a single day ahead forecaster.
Residual analysis shows that beyond 500 customers, gaussian residuals is a reasonable model.
We then show the forecaster has uncorrelated errors.
\end{abstract}

\section{Statistical Modeling of Protective Loads}
\label{section-Load-Modeling}

The model for the distribution system used in \cite{Sevlian2014A_Outage} and \cite{Sevlian2014B_Outage} is of a tree $\mathcal{T}$ where vertices in the network are loads while edges of the network are protective relays which can disconnect loads from the network.
To determine a statistical model of the loads for use in simulation we rely on the Pacific Northwest National Laboratories Taxonomy Feeders \cite{PNNL2008}.
For each one of the feeders, the individual loads were aggregated around the respective protective device.
For this analysis, fuses and switches are considered.
We analyze all 23 feeders and perform aggregate analysis based on the combined load information.

\begin{figure}[h]
\centering
\subfigure[][] { 
	\includegraphics[width=0.22\textwidth, height=0.22\textwidth]{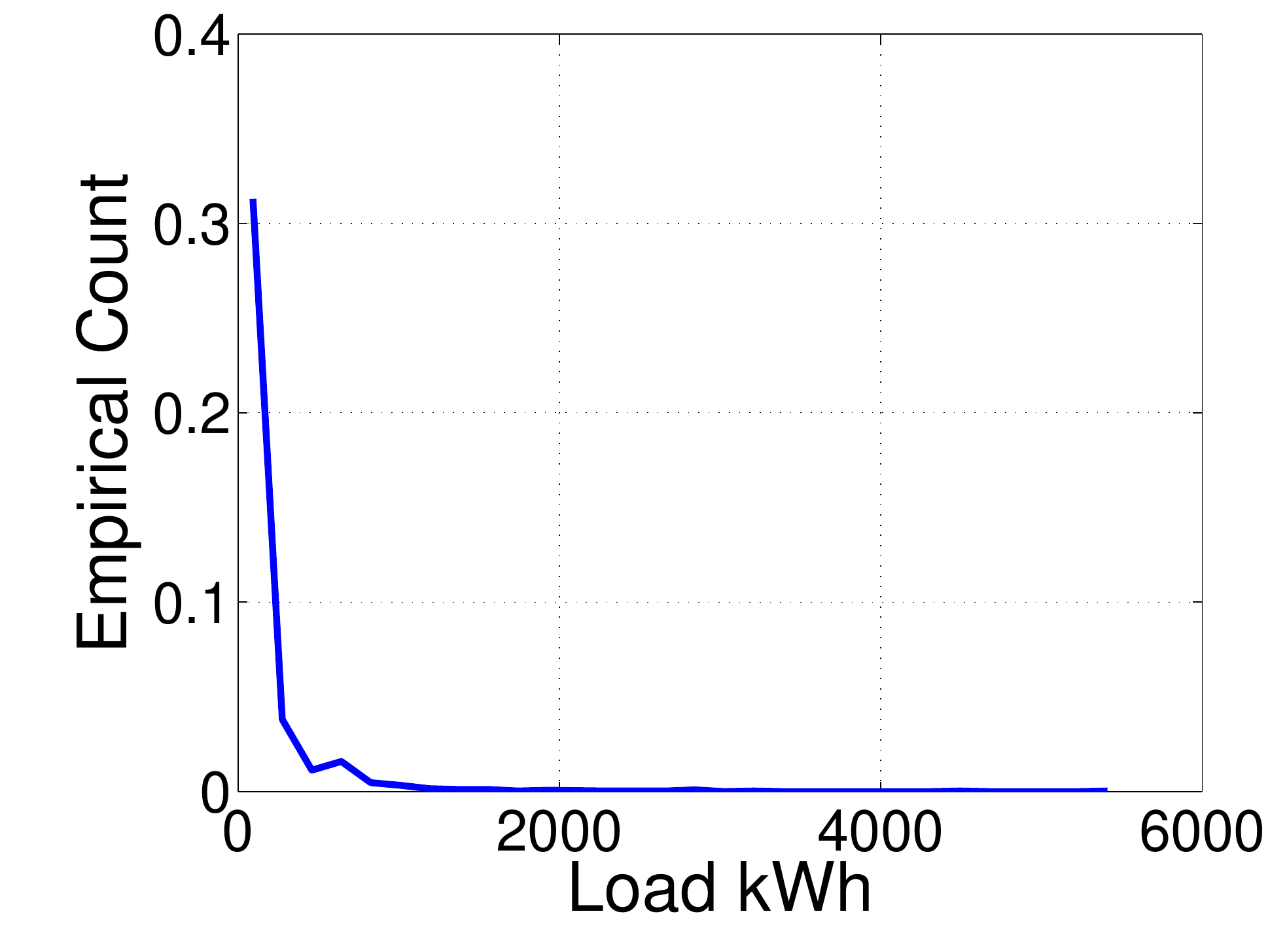}
	\label{fig:empirical_histogram}
}
\hspace{-5mm}
\subfigure[][] { 
	\includegraphics[width=0.22\textwidth, height=0.22\textwidth]{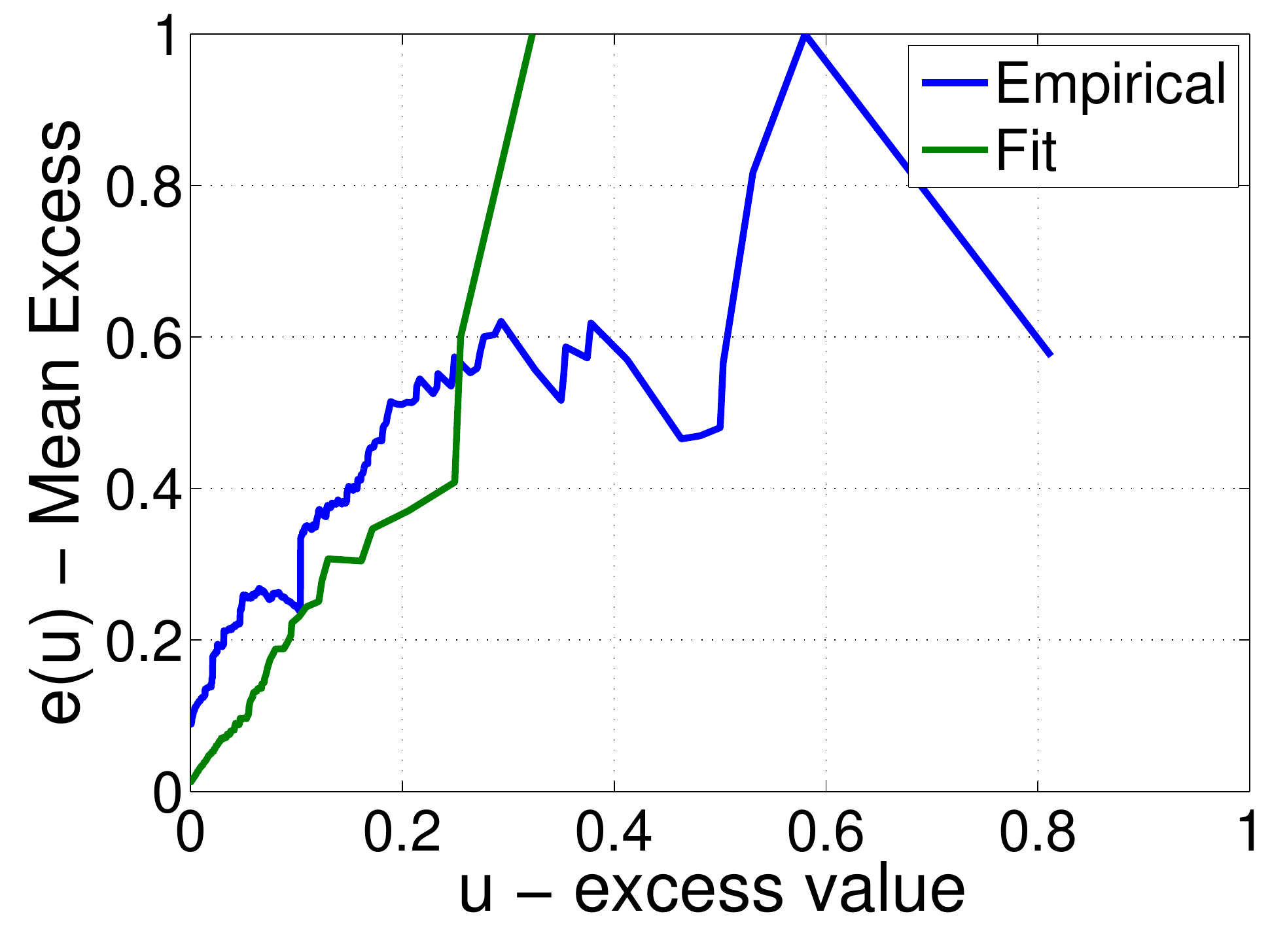}
	\label{fig:mean_excess_function}
}
\hspace{-5mm}
\subfigure[][] { 
	\includegraphics[width=0.23\textwidth, height=0.22\textwidth]{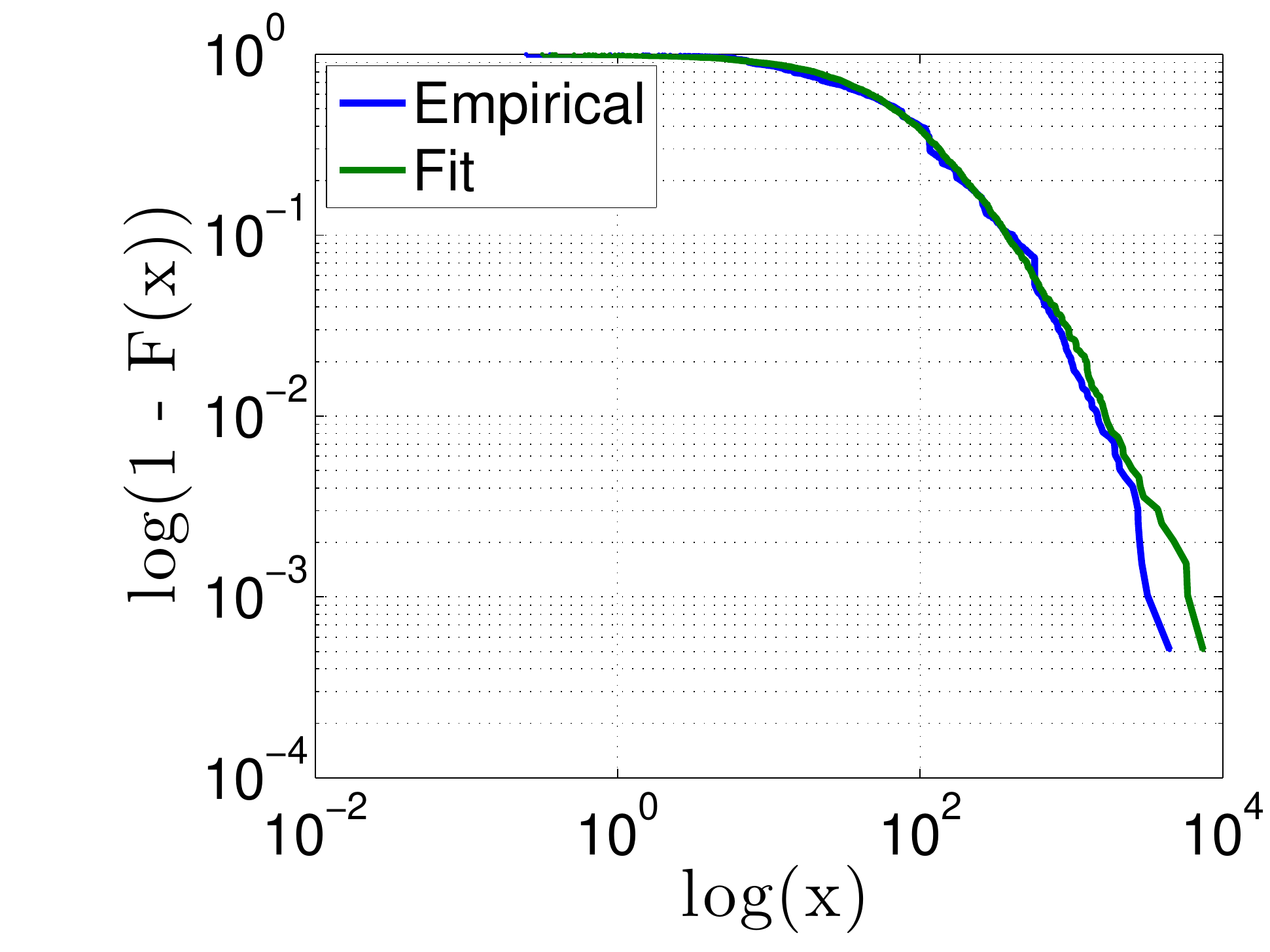}
	\label{fig:log_survival}
}
\hspace{-5mm}
\subfigure[][] { 
	\includegraphics[width=0.23\textwidth, height=0.21\textwidth]{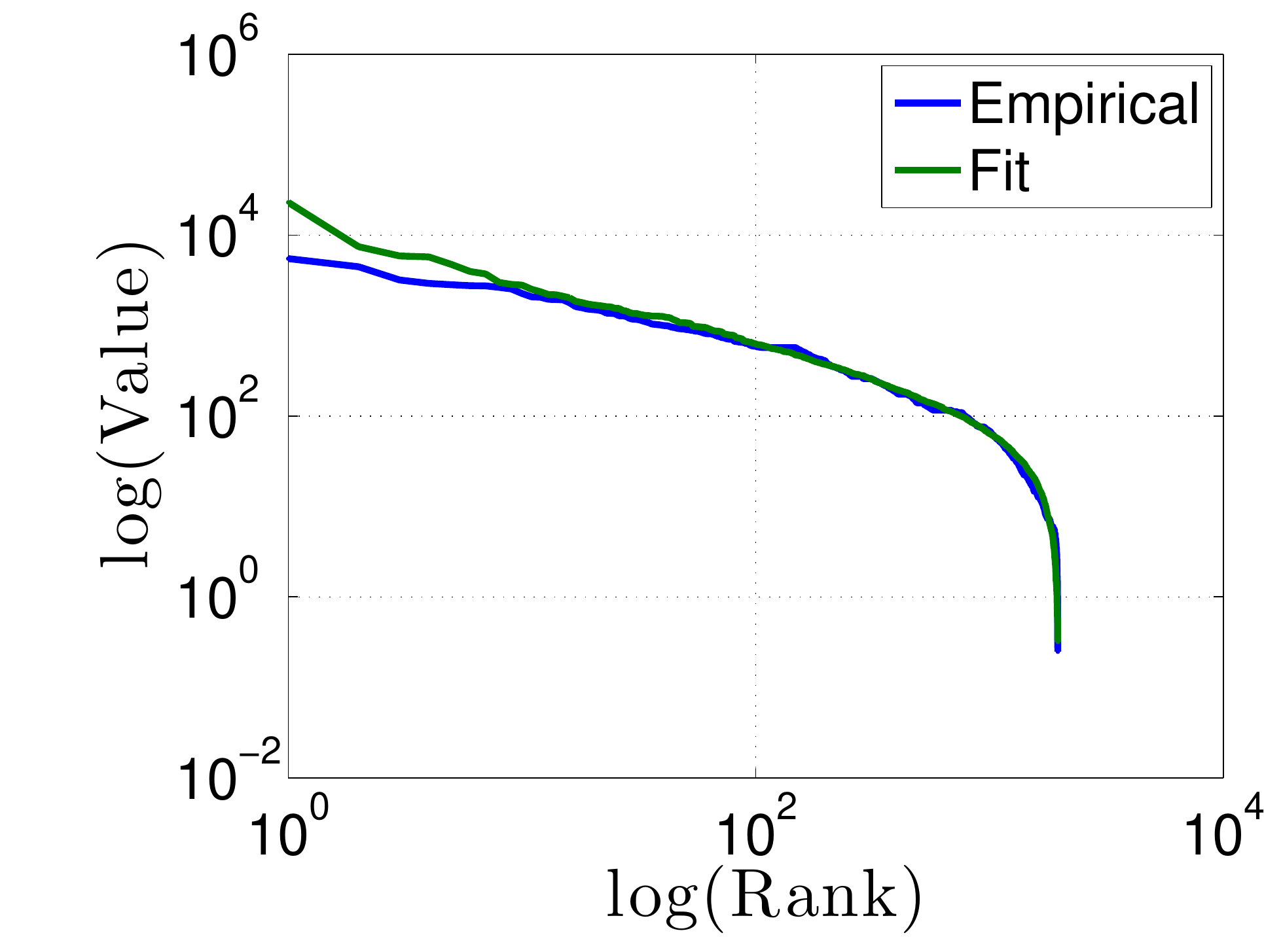}
	\label{fig:log_rank_plot}
}
\caption{ \ref{fig:empirical_histogram} Empirical histogram of all loads collected from 23 feeders.  
An individual load is the grouping of all loads that  can be disconnected from the feeder root using either switches or fuses.
The empirical distribution is clearly long tailed.
Proposed model is generalized Pareto model as per Eq. \eqref{eq:general-pareto-distribution}.
Model fit given in Table \ref{tab:general_pareto_fit}.
\ref{fig:mean_excess_function} Validation of heavy tail distribution mean excess function $e(u)$ for the load data as well as a randomly generated dataset using  parameter values from Table \ref{tab:general_pareto_fit}.
\ref{fig:log_survival} Log survival function for data as well as randomly generated data using fit shows close agreement.
\ref{fig:log_rank_plot} Zipf or log rank plot for data as well as randomly generated data using fit shows close agreement.
}

\label{fig:unv-M1-log-log}
\end{figure}

An empirical histogram of the load values is given in Figure \ref{fig:empirical_histogram}.
We propose that the loads at each node in the network follow a generalized Pareto distribution which is of the form.

\begin{align}
f(x; \kappa, \sigma, \theta ) = \left( \frac{1}{\sigma} \right)
 \left( 1 + \kappa \frac{x - \theta }{ \sigma} \right) ^{- (\kappa-1)/\kappa } \label{eq:general-pareto-distribution}
\end{align}

\vspace{+5mm}
\begin{table}[h]
\hspace{+5mm}
\label{tab:general_pareto_fit}
\begin{tabular}{@{}ccccc@{}}
\toprule
  \cmidrule{1-5}  
Parameter     &  & $\theta$ & $\kappa$ & $\sigma$      \\ 
Value             &  & 0.25       &  0.58 (0.51 0.65) &     74.28 (68.53 80.50)  \\ 
\bottomrule
\end{tabular}
\caption{Maximum Likelihood Parameter estimate with 95 $\%$ Confidence interval for General Pareto distributino e.q. \eqref{eq:general-pareto-distribution}. } 
\end{table}

This is determined by maximum likelihood estimate of the model parameters.
The estimated values with $95\%$ confidence intervals for the parameters are shown in Table \ref{tab:general_pareto_fit}.

In general, visual procedures such as this are the more useful way of diagnosing the particular nature of the data, before any fitting is done \cite{ParetoRef2013}.
To validate the heavy tail distribution we consider a mean excess function (Figure \ref{fig:mean_excess_function}), log survival function (Figure \ref{fig:log_survival}) and a zipf plot (Figure \ref{fig:log_rank_plot}) of both the load data and randomly sampled distribution following the fit parameters of Table \ref{tab:general_pareto_fit}.

The mean excess function computes $e(u) = \mathbf{E}\left[X-u | X > u \right]$ which for Pareto distribution is linearly increasing.
Figure \ref{fig:mean_excess_function} shows that the actual load data as well as the sampled data show similar increase excess functions.

The Zipf plot will produce on the x axis $\{ \log 1, \hdots, \log N \}$ and on the y axis $\{ \log x_{[1]}, \hdots, \log x_{[N]} \}$, where $x_{[k]}$ is the $k^{th}$ largest element of the dataset.
In the case of a Pareto distribution the Zipf plot is lineally decreasing with slope equal to the exponent of the power law.
Figure \ref{fig:log_rank_plot} shows that this is not the case.  
However, the rank plot is linear for values $> 10^1$, which indicates that the tail of the distribution can be approximated as a Pareto distribution.

\section{Distribution Level Load Forecasting}
\label{section-Distribution-Level-Load-Forecasting}

The forecasting model is for day ahead consumption.  
Let $x_{d} \in \mathbb{R}^{24}$ be the aggregate daily consumption for days $d = \{1, \hdots, D \}$.
Given previous consumption information $\mathcal{X}_d = \{ x_{1}, \hdots, x_{d} \}$ and daily temperature forecasts $T_d = \{t_1, \hdots, t_{d+1} \}$ the forecaster will output the next day's consumption profile $\hat{x}_{d+1}$.

The forecaster works by predicting the daily total consumption $\hat{p}_d \in \mathbb{R}$ and normalized daily shape pattern $\hat{u}_d \in \mathbb{R}^{24}$ separately.
The final prediction $\hat{x}_{d+1} = \hat{p}_{d+1} \hat{u}_{d+1}$ is the product of each individual forecast.

\textit{Total Power Forecaster:} 
An autoregressive moving average with exogenous input (armax) model is used to forecast the total consumption which is of the form 
\begin{align}
\hat{p}_{d+1} = \sum_{k=d+1-K}^{d} a_{k} p_{k} + \sum_{r=d+1-K}^{d+1} b_{r} t_{r} \\
\end{align}

The exogenous input $t_r \in \mathbb{R}$ is the daily mean temperature.
The parameters $a_k,b_r \in \mathbb{R}$ are determined by least squares regression.
A cross validation stage is used to estimate the proper model size $K$.

\textit{Shape Forecaster:} 
A vector armax method is used to forecast the daily shape profile.  
The model is of the form

\begin{align}
\hat{u}_{d+1} = \sum_{k=d+1-K}^{d} C_{k} u_{k} + \sum_{r=d+1-K}^{d+1} h_{r} t_{r} \\
\end{align}

The exogenous input $t_r \in \mathbb{R}^{24}$ contains the mean temperature for each hour.
The parameters $C_{k},h_{r} \in \mathbb{R}^{24 \times 24}$ are real matrices.
These parameters are determined by linear regression given the training data using a least squares formulation.
The model size $K$ is determined in a cross validation stage.

\section{Load Data and Model Fitting}

\subsection{Forecast Error On Load}
\label{subsection-Forecast Error On Load}
~~~~~~ \\
\textit{Coefficient of Variation Scaling:} It has been shown recently \cite{Sevlian2014_Short}, \cite{Sevlian2014_Model}, that load forecasting at the small loads generally follows an empirical scaling law.
In \cite{Sevlian2013_Value} the authors propose using aggregation level (in terms of number of customers, or meal kWh) as a system parameter in distribution system applications.
These studies have focused on the very short term forecasts of one hour to multiple hours ahead.
In all this work, the authors propose the following model to represent the forecasting scaling law.

\begin{align}
\overline{\textrm{CV}}(W) &= \textbf{E}[\textrm{CV}(x_A, \hat{x}_A)\left| W_A = W \right.]\nonumber\\ 
&=  \sqrt{ \frac{\beta_{0}}{W^p} + \beta_1 } \ (\%).  \label{eq:CV-with-load}
\end{align}

For the simulation study, we focus on a day ahead forecast of the load values.
Although other time horizons are also possible, day ahead is convenient for analysis.
We focus on the intuitive model with $p=1$ leading to the following fit: $\overline{\textrm{CV}}(W) =  \sqrt{ \frac{\beta_{0}}{W} + \beta_1 } \ (\%)$. 

\begin{figure}[h]
\centering
\includegraphics[width=0.35\textwidth, height=0.28\textwidth]{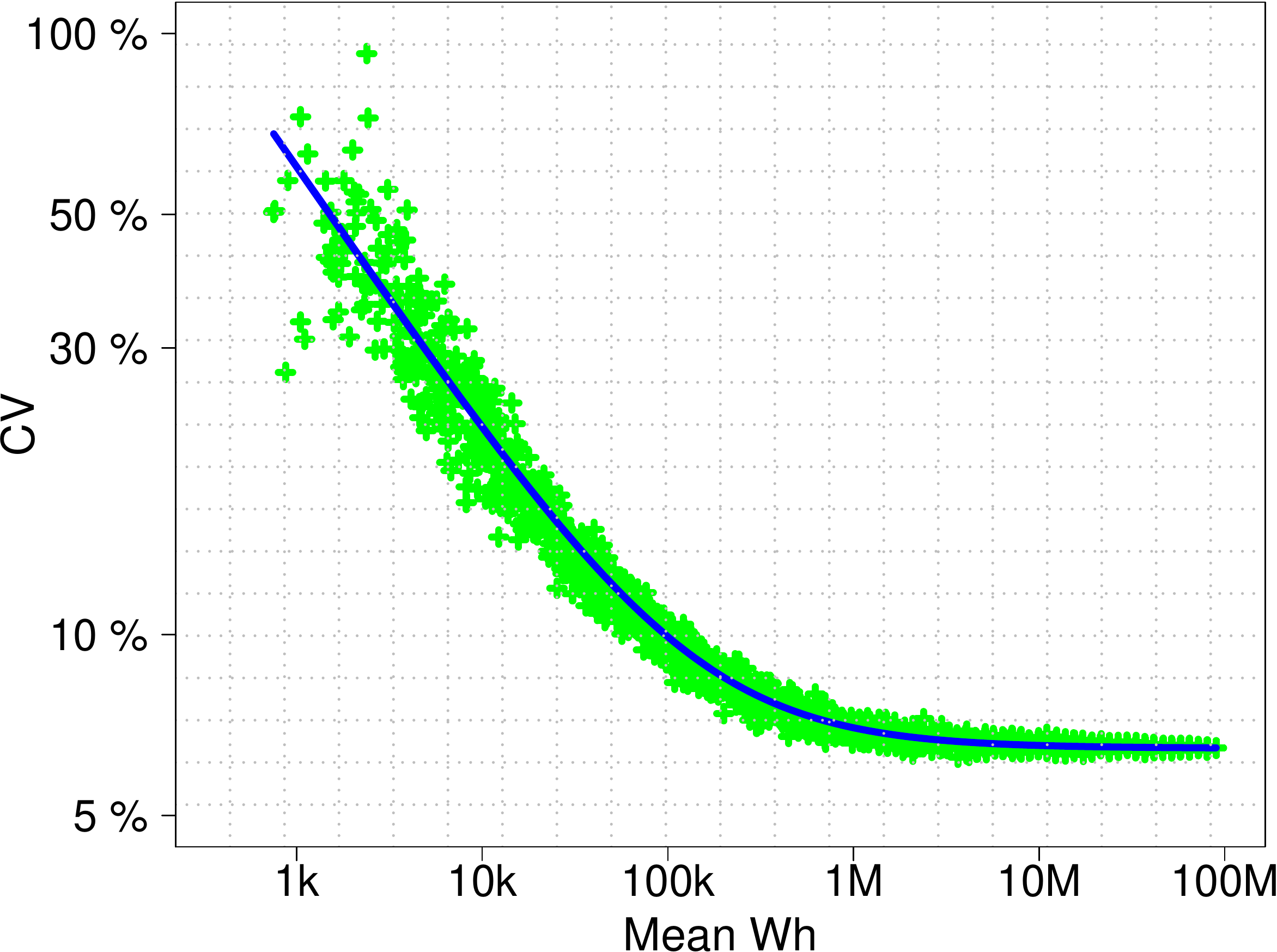}
\caption{ Aggregation error curve for day ahead forecasting of $PG \& E$ loads.
Non-linear Least square model parameters $\alpha_0 = 3562$ and $\alpha_1 = 41.9$ with $p = 1$.
This leads to a critical load $W^{\star}$ of 85 kWh and irreducible error of $6.47 \%$}
\label{fig:day_ahead_aggregation_error} 
\end{figure}

\begin{figure}[h]
\hspace{-5mm}

\subfigure[][] { 
	\includegraphics[width=0.23\textwidth, height=0.19\textwidth]{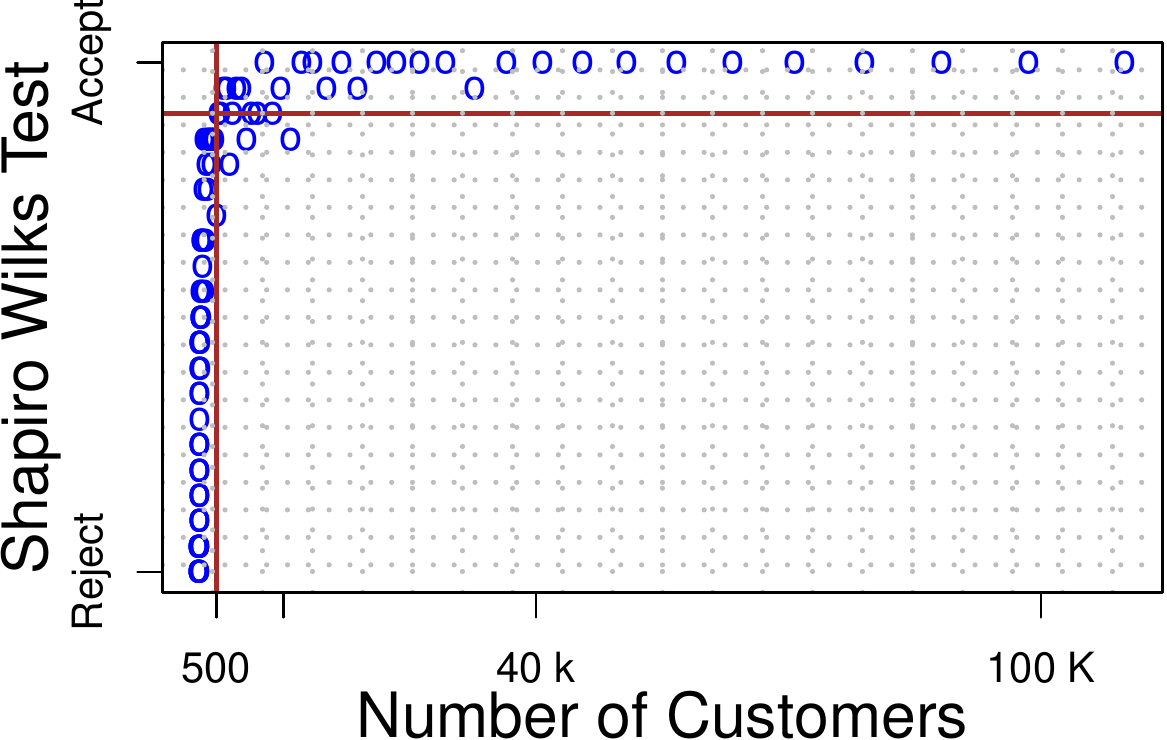}
	\label{fig:day_ahead_shapiro_wilks}
}
\hspace{-3mm}
\subfigure[][] { 
	\includegraphics[width=0.23\textwidth, height=0.18\textwidth]{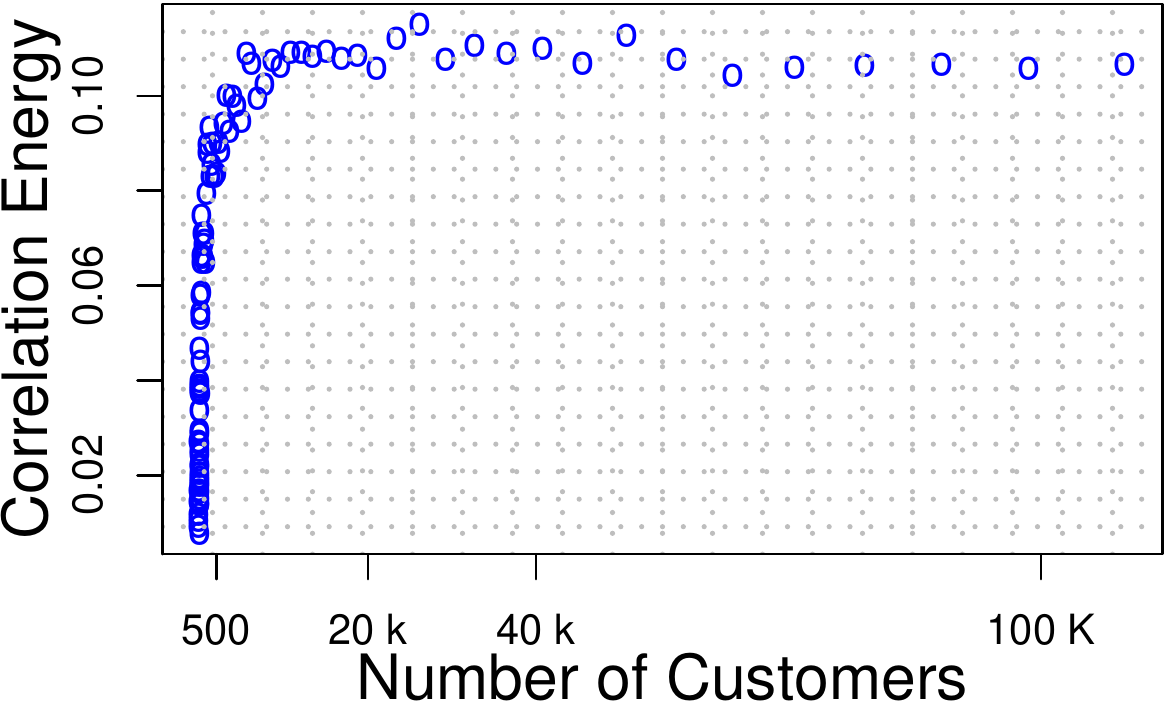}
	\label{fig:day_ahead_correlation_energy}
}
\caption{ Aggregation statistics for mean load and forecasts. 
 \ref{fig:day_ahead_shapiro_wilks} shows normality of forecast errors beyond 500 users.  This work assumes normality below this point. 
 \ref{fig:day_ahead_correlation_energy} shows correlation energy outside of first component for varying aggregation levels.  Work indicates uncorrelated error reasonable model. }
\label{fig:unv-M1-log-log}
\end{figure}

Using the model in Section \ref{section-Distribution-Level-Load-Forecasting} the aggregation error curve for randomly generated loads in climate zone 13 of california are shown in Figure \ref{fig:day_ahead_aggregation_error}.
Also, fitting the model proposed in \cite{Sevlian2014_Short} and \cite{Sevlian2014_Model} we have the following approximate model $\overline{\textrm{CV}}(W) =  \sqrt{ \frac{3562 }{W} + 41.9}$.
This leads to a irreducible error of $6.47 \%$ for a full day ahead forecaster and a critical load of $W^{\star} = 85$ kWh, or around 100 homes.  
However, from Figure \ref{fig:day_ahead_aggregation_error}, it appears as though forecast performance flattens our after 1 MWh of load.
Therefore, using the model in eq. \eqref{eq:CV-with-load} for any load less than 1 MWh is useful in understanding the dependence of load size to aggregation.
For the case studies in this work, all node load values are less than 1 MWh thereby necessitating the use of this model.

\textit{Residual Gaussianity:}

This work assumes normality of the forecast error.  
The central limit theorem indicates that a larger aggregation levels, the residuals should look more gaussian.
We test this by running a day ahead forecaster on each aggregation level and performing a shapiro wilks test on the residual time series.
Each each aggregation level, there are 20 time series, so we evaluate the percentage of tests that pass the test given for a level of $95 \%$.
Ideally, the test should pass $95\%$ under a normally distributed residual.

Figure \ref{fig:day_ahead_shapiro_wilks} shows the Shapiro Wilks test for varying aggregation levels. 
The horizontal line indicates the $95\%$ acceptance line.
The point at which the curve crosses the $95\%$ line is 500 users.
We assume in this work that the residuals are normal at all levels of aggregation considered. 
We do this since, forecasters can be tuned to have normal residuals, even though they are not considered here

~~~~~~~~\\
\textit{Residual Correlation:}

We also assume iid forecast errors in this work.
Figure \ref{fig:day_ahead_correlation_energy} shows the amount of energy outside of the first tap of the autocorrelation function.  
More specifically, given the forecast residual $e[n]$, the sample normalized autocorrelation computes the following:
\begin{align}
\rho[m] = \frac{ \sum_{n} e[n+m]e[n] }{ \sum_{n} e[n]^2}
\end{align}

Given an autocorrelation function $\rho[m]$ we use the total statistically significant values of $\rho[m]$ where $m \neq 0$.
So the 'correlation energy' for a time series is computed as:
\begin{align}
\gamma = \frac{ \sum_{m\neq0} | \rho[m] | }{ \sum_{m} | \rho[m] | }
\end{align}

Figure \ref{fig:day_ahead_correlation_energy} shows the mean value of $\gamma$ for each aggregation level. 
The figure shows that there is at most $10 \%$ of the total energy in taps beyond the first one. 
Although ideally this value should be zero, it leads us to conclude that the iid error model is reasonable.

\bibliographystyle{plain}
\bibliography{day_ahead_bib}
\end{document}